\documentclass[fleqn,twoside]{article}
\usepackage{amsmath,espcrc2,graphicx}

\begin{document}

\pagestyle{empty}

\begin{flushleft}
\Large{SAGA-HE-208-04     \hfill July 31, 2004}
\end{flushleft}
\vspace{2.2cm}
 
\begin{center}
\LARGE{\bf Nuclear corrections of \\ parton distribution functions} \\
\vspace{1.2cm}
\Large{ M. Hirai $^{(a)}$, S. Kumano $^{(b), *}$, and T.-H. Nagai $^{(b)}$}  \\
\vspace{0.4cm}
\Large
{(a) Institute of Particle and Nuclear Studies, KEK \\

     1-1, Ooho, Tsukuba, Ibaraki, 305-0801, Japan} \\
     \vspace{0.5cm}   
{(b) Department of Physics, Saga University \\
     Saga, 840-8502, Japan} \\
\vspace{2.2cm}
\Large{Talk at the Third International Workshop on \\
       Neutrino-Nucleus Interactions in the Few GeV Region (NuInt04)}

\vspace{0.6cm}

{Gran Sasso, Italy,  March 17 - 21, 2004} \\

\vspace{0.5cm}
{(talk on March 17)}  \\
\end{center}
\vspace{2.0cm}
\noindent{\rule{6.0cm}{0.1mm}} \\
\vspace{-0.3cm}
\normalsize

\noindent
{* URL: http://hs.phys.saga-u.ac.jp.}  \\

\vspace{+0.5cm}
\hfill {\large to be published in Nucl. Phys. B Supplements}
\vfill\eject
\setcounter{page}{1}
\pagestyle{headings}

\title{Nuclear corrections of parton distribution functions}
\author{M. Hirai
     \address{Institute of Particle and Nuclear Studies, KEK,
              1-1, Ooho, Tsukuba, Ibaraki, 305-0801, Japan}
     \thanks{mhirai@rarfaxp.riken.go.jp},
     S. Kumano
     \address[saga]{Department of Physics, Saga University,
              Saga, 840-8502, Japan}
     \thanks{kumanos@cc.saga-u.ac.jp, http://hs.phys.saga-u.ac.jp},
     and T.-H. Nagai
     \addressmark[saga]
     \thanks{03sm27@edu.cc.saga-u.ac.jp}
     }


\begin{abstract}
We report global analysis results of experimental data for
nuclear structure-function ratios $F_2^A/F_2^{A'}$ and proton-nucleus
Drell-Yan cross-section ratios $\sigma_{DY}^{pA}/\sigma_{DY}^{pA'}$
in order to determine optimum parton distribution functions (PDFs) 
in nuclei. An important point of this analysis is to show uncertainties of
the distributions by the Hessian method. The results indicate
that the uncertainties are large for gluon distributions
in the whole $x$ region and for antiquark distributions at $x>0.2$. 
We provide a code for calculating any nuclear PDFs at given $x$ and $Q^2$
for general users. They can be used for calculating high-energy
nuclear reactions including neutrino-nucleus interactions,
which are discussed at this workshop.
\vspace{1pc}
\end{abstract}
\maketitle

\section{Introduction}

Although parton distribution functions (PDFs) in the nucleon are now
determined relatively well in the wide range of $x$, their nuclear
modifications are not determined accurately. There are experimental
measurements about nuclear effects on the PDFs, for example
nuclear $F_2$ data, so that their gross properties are known.
However, the details are not still determined because available
experimental data exist for a limited number of high-energy
nuclear reactions.

On the other hand, accurate nuclear parton distribution functions (NPDFs)
\cite{npdf01,npdf04,other-npdfs,related} are needed for describing
any high-energy nuclear reactions such as heavy-ion and neutrino reactions.
The major purpose of this workshop is to discuss nuclear effects
on neutrino reactions \cite{nuint04}. Since neutrino-oscillation
experiments \cite{nu-osci} become more and more accurate, it is now
important to take nuclear medium effects into account.

Current oscillation experiments are done in a medium-energy
region, so that there are various nuclear effects which contribute to 
the neutrino reactions. Among them, we investigate nuclear medium
effects on the PDFs. Of course, the parton distributions are supposed
to be used in the deep inelastic (DIS) region, so that an extension to
a smaller $Q^2$ region becomes important in order to use them
for the present neutrino reactions. 

In this paper, we report our recent studies on the NPDFs by analyzing
the nuclear data on the structure-function ratios $F_2^A/F_2^{A'}$ and
Drell-Yan cross-section ratios $\sigma_{DY}^{pA}/\sigma_{DY}^{pA'}$
\cite{npdf04}. We had already reported the NPDF studies on the first version
\cite{npdf01} at the NuInt01 workshop \cite{nuint01} and
preliminary studies after the first version at the NuInt02 \cite{nuint02}.

The most important point in Ref. \cite{npdf04} is that uncertainties
of the NPDFs are estimated by the Hessian method. For calculating
any nuclear reactions, it is especially important to show 
the uncertainties which come from the PDFs. 
Furthermore, Drell-Yan and HERMES data are added into the data set and
the charm-quark distributions are included in the new analysis.

This paper consists of the following.
In section \ref{method}, a method is explained for analyzing 
$F_2$ and Drell-Yan data in order to obtain the optimum NPDFs.
Analysis results are shown in section \ref{results}.
We provide a code for calculating the NPDFs at given $x$ and $Q^2$,
and it is explained in section \ref{code}.
The results are summarized in section \ref{summary}.

\section{Analysis method}
\label{method}

The parton distribution functions are expressed in general by two variables,
$x$ and $Q^2$. In lepton-nucleus scattering, the $Q^2$ is defined by
the momentum transfer $q$: $Q^2=-q^2$, and the Bjorken scaling variable $x$
is  given by $x=Q^2/(2 M \nu)$ with the nucleon mass $M$ and the energy
transfer $\nu$.
The NPDFs are provided by an analytical form at a fixed $Q^2$ point
($Q_0^2$). Practically, a NPDF could be expressed by the corresponding 
nucleonic distribution multiplied by a function $w_i$:
\begin{equation}
f_i^A (x, Q_0^2) = w_i(x,A,Z) \, f_i (x, Q_0^2).
\label{eqn:paramet}
\end{equation}
We call $w_i$ a weight function, which indicates nuclear modification.
The function is expressed by a number of parameters:
\begin{align}
w_i(x,A,Z) & = 1+\left( 1 - \frac{1}{A^{1/3}} \right) 
\nonumber \\
  &  \times    \frac{a_i(A,Z) +b_i x+c_i x^2 +d_i x^3}{(1-x)^{\beta_i}}.
\label{eqn:wi}
\end{align}

The parameters are determined by a $\chi^2$ analysis with
experimental data on structure functions $F_2^A$ and
Drell-Yan cross sections $\sigma_{DY}^{pA}$.
Because of baryon-number, charge, and momentum conservations,
three parameters can be fixed.
The cubic functional form of the numerator is motivated by 
the $x$ dependence of typical data for $F_2^A/F_2^D$,
and the factor $1/(1-x)^{\beta_i}$ is to reproduce
the Fermi-motion part at large $x$. 
The nuclear dependence $1 - 1/A^{1/3}$ is introduced in
Ref. \cite{sd} simply by considering nuclear volume and surface
contributions to cross sections.
Because different physics mechanisms contribute
in each $x$ region, the overall $1/A^{1/3}$ dependence would be
too simple to describe the nuclear modifications. 
For the NPDFs $f_i^A$, we take $u_v^A$, $d_v^A$, $\bar q^A$, and $g^A$
by assuming flavor symmetric antiquark distributions although
they are not symmetric in the nucleon \cite{flavor3}.

The parameters are determined by a $\chi^2$ analysis with experimental
values. The $\chi^2$ is given by 
\begin{equation}
\chi^2 = \sum_j \frac{(R_{j}^{data}-R_{j}^{theo})^2}
                     {(\sigma_j^{data})^2},
\label{eqn:chi2}
\end{equation}
where $\sigma_j^{data}$ is an experimental error, and
$R_j$ indicates a ratio, $F_2^A/F_2^{A'}$ or
$\sigma_{DY}^{pA}/\sigma_{DY}^{pA'}$.
Leading-order expressions are used in the theoretical calculations.

The actual calculation is done by running the subroutine {\tt MINUIT}.
The subroutine also produces a Hessian matrix $H$ which has information
on parameter errors. Using the matrix, we can
calculate the uncertainty of a NPDF:
\begin{align}
        [\delta f^A(x)]^2 = \Delta \chi^2  &  \sum_{i,j} 
          \left( \frac{\partial f^A(x,\xi)}{\partial \xi_i} 
                             \right)_{\xi=\hat\xi}
\nonumber \\
   & \times       H_{ij}^{-1}
          \left( \frac{\partial f^A(x,\xi)}{\partial \xi_j} 
                             \right)_{\xi=\hat\xi}
\, ,
        \label{eq:dnpdf}
\end{align}
where $\xi_i$ is a parameter, $\hat\xi$ indicates the optimum
set of the parameters, and $\delta f^A(x)$ is the uncertainty
of the nuclear PDF $f^A(x)$. 
The $\Delta \chi^2$ value determines a confidence region and
it depends on the number of parameters. 

\begin{figure}[b]
\vspace{-0.7cm}
\begin{center}
     \includegraphics[width=0.41\textwidth]{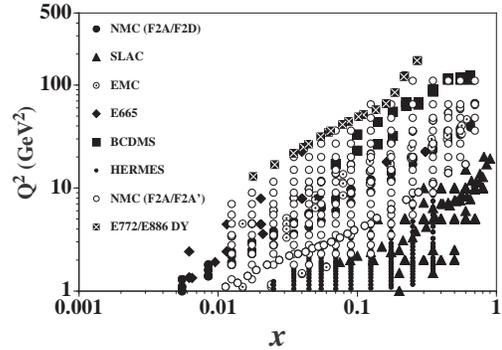}
\end{center}
\vspace{-1.2cm}
\caption{Kinematical region is shown for experimental data.}
\vspace{-0.7cm}
\label{fig:xq2}
\end{figure}

The kinematical region of the experimental data for
$F_2^A$ and $\sigma_{DY}^{pA}$ is shown in Fig. \ref{fig:xq2}.
In comparison with the nucleon data, the $x$ range is still limited. 
Namely, the small-$x$ ($x=10^{-5} - 10^{-3}$)
data are not taken. The Drell-Yan data are taken in the large $Q^2$ region.
There are 606 data points for the $F_2^A/F_2^D$ type, 293
for $F_2^A/F_2^{A'}$ ($A' \ne D$),
and 52 for the Drell-Yan. The total number of data is 951.
These data are taken for the targets:
deuteron (D), helium-4 ($^4$He), lithium (Li), beryllium (Be), carbon (C),
nitrogen (N), aluminum (Al), calcium (Ca), iron (Fe), copper (Cu),
krypton (Kr), silver (Ag), tin (Sn), xenon (Xe), tungsten (W), gold (Au),
and lead (Pb).

\section{Results}
\label{results}

\subsection{Comparison with experimental data}

The NPDFs in Eqs. (\ref{eqn:paramet}) and (\ref{eqn:wi}) are given
at $Q_0^2$=1 GeV$^2$, and they are evolved to the experimental $Q^2$ points
for calculating the total $\chi^2$ in Eq. (\ref{eqn:chi2}).
The optimum parameters are obtained by minimizing
$\chi^2$. The minimum $\chi^2$ becomes 1489.8 for the 951 data.
The uncertainties of the NPDFs are estimated with $\Delta \chi^2$=10.427
in order to show the one-$\sigma$-error range.
It is chosen by considering that the number of the parameters is nine
\cite{npdf04,aac03}.

\begin{table}[b!]
\vspace{-0.7cm}
\caption{Each $\chi^2$ contribution.}
\label{tab:chi2}
\footnotesize
\begin{tabular*}{\hsize}
{c@{\extracolsep{0ptplus1fil}}c@{\extracolsep{0ptplus1fil}}c}
\hline
nucleus  & \# of data & $\chi^2$  \\
\hline
$^4$He/D &   \   35    &  \     56.0         \\
Li/D     &   \   17    &  \     88.7         \\ 
Be/D     &   \   17    &  \     44.1         \\
C/D      &   \   43    &       130.8         \\
N/D      &   \  162    &       136.9         \\
Al/D     &   \   35    &  \     43.1         \\
Ca/D     &   \   33    &  \     42.0         \\
Fe/D     &   \   57    &  \     95.7         \\
Cu/D     &   \   19    &  \     11.8         \\
Kr/D     &      144    &       126.9         \\
Ag/D     &   \ \  7    &  \     12.8         \\
Sn/D     &   \ \  8    &  \     14.6         \\
Xe/D     &   \ \  5    &  \ \    2.0         \\
Au/D     &   \   19    &  \     61.6         \\
Pb/D     &   \ \  5    &  \ \    5.6         \\	
\hline
$F_2^A/F_2^D$ total
         &      606    &       872.8         \\
\hline
Be/C     &   \   15    &  \     16.1         \\
Al/C     &   \   15    &  \ \    6.1         \\
Ca/C     &   \   39    &  \     36.5         \\
Fe/C     &   \   15    &  \     10.3         \\
Sn/C     &      146    &       257.3         \\
Pb/C     &   \   15    &  \     25.3         \\
C/Li     &   \   24    &  \     78.1         \\
Ca/Li    &   \   24    &       107.7         \\
\hline
$F_2^{A_1}/F_2^{A_2}$ total 
         &      293    &       537.4         \\
\hline
C/D      &   \ \  9    &  \ \    9.8         \\
Ca/D     &   \ \  9    &  \ \    7.2         \\
Fe/D     &   \ \  9    &  \ \    8.1         \\
W/D      &   \ \  9    &  \     18.3         \\
Fe/Be    &   \ \  8    &  \ \    6.5         \\
W/Be     &   \ \  8    &  \     29.6         \\
\hline
Drell-Yan total  
         &   \   52    &  \     79.6         \\
\hline\hline
total    &      951    &       1489.8 \      \\
\hline
\end{tabular*}
\end{table}

Each $\chi^2$ contribution is listed in Table \ref{tab:chi2}.
There is a tendency that the $\chi^2$ values are large for small nuclei.
The $Li/D$, $Be/D$, and $C/D$ ratios have $\chi^2$ values,
88.7, 44.1, and 130.8, for the number of data points, 17, 17, and 43
points, respectively. Medium and large size nuclei are generally 
well reproduced except for the $Sn/C$ and $Ca/Li$ ratios.
The Drell-Yan data are also well explained except for the $W/Be$ ratios.

\begin{figure}[b!]
\vspace{-0.5cm}
\begin{center}
     \includegraphics[width=0.30\textwidth]{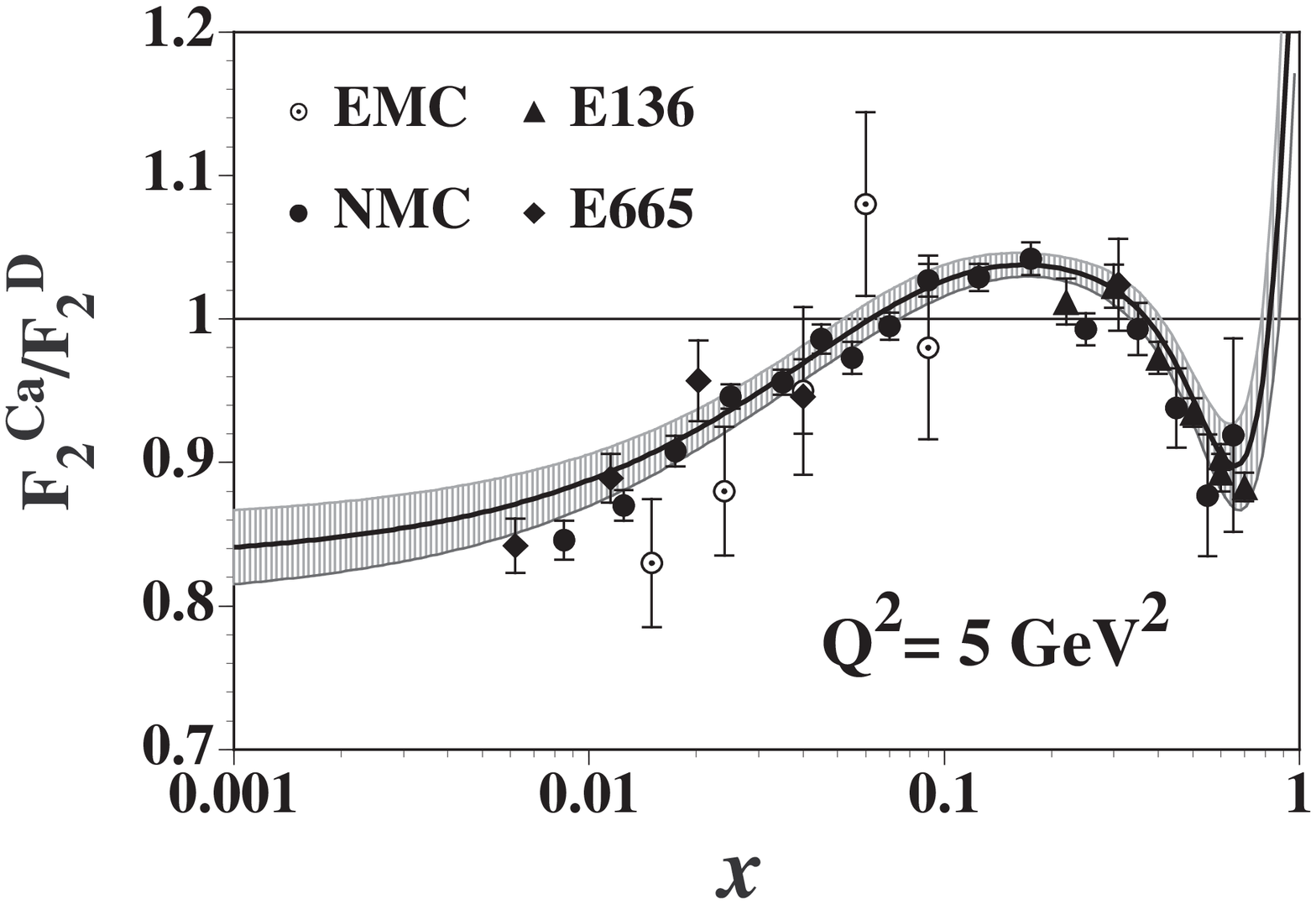}
\end{center}
\vspace{-1.3cm}
\caption{Comparison with the $F_2^{Ca}/F_2^D$ data.}
\label{fig:f2cad}
\vspace{0.1cm}
\begin{center}
     \includegraphics[width=0.30\textwidth]{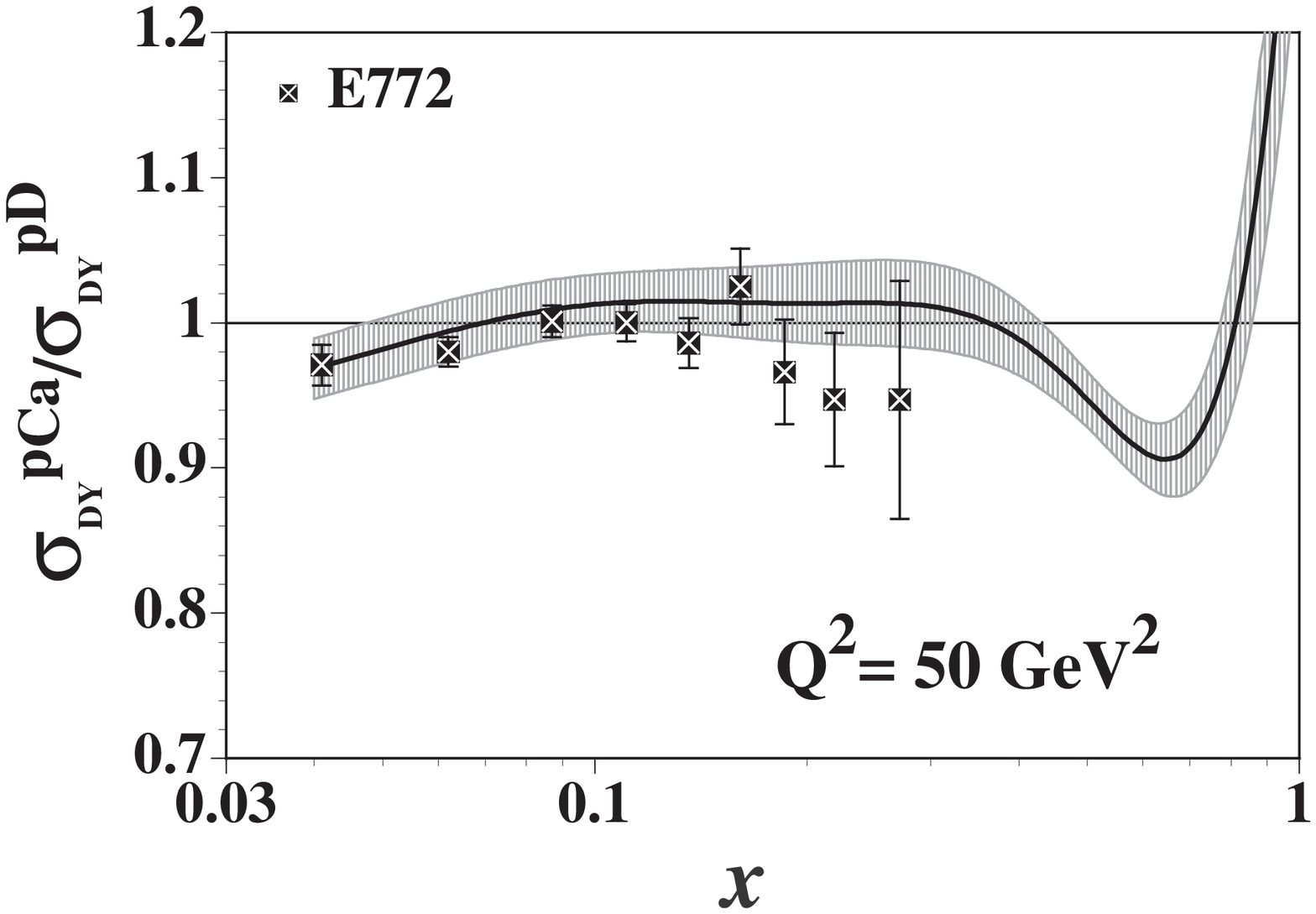}
\end{center}
\vspace{-1.3cm}
\caption{Comparison with the $\sigma_{DY}^{pCa}/\sigma_{DY}^{pD}$ data.}
\label{fig:dycad}
\vspace{-0.5cm}
\end{figure}

Typical results are shown in Figs. \ref{fig:f2cad} and \ref{fig:dycad}.
In Fig. \ref{fig:f2cad}, the $F_2^{Ca}/F_2^D$ data are compared with
the fit result at $Q^2$=5 GeV$^2$. The shaded area is the uncertainty
range due to the NPDF uncertainties at $Q^2$=5 GeV$^2$. One should
note that the experimental data are taken at various $Q^2$ points which
are not equal to 5 GeV$^2$. Therefore, the curve cannot be, strictly
speaking, compared with the data. However, considering that
the $Q^2$ dependence is small in general, we find that
the data are reproduced well by the parametrization. 

\begin{figure}[b!]
\begin{center}
     \includegraphics[width=0.30\textwidth]{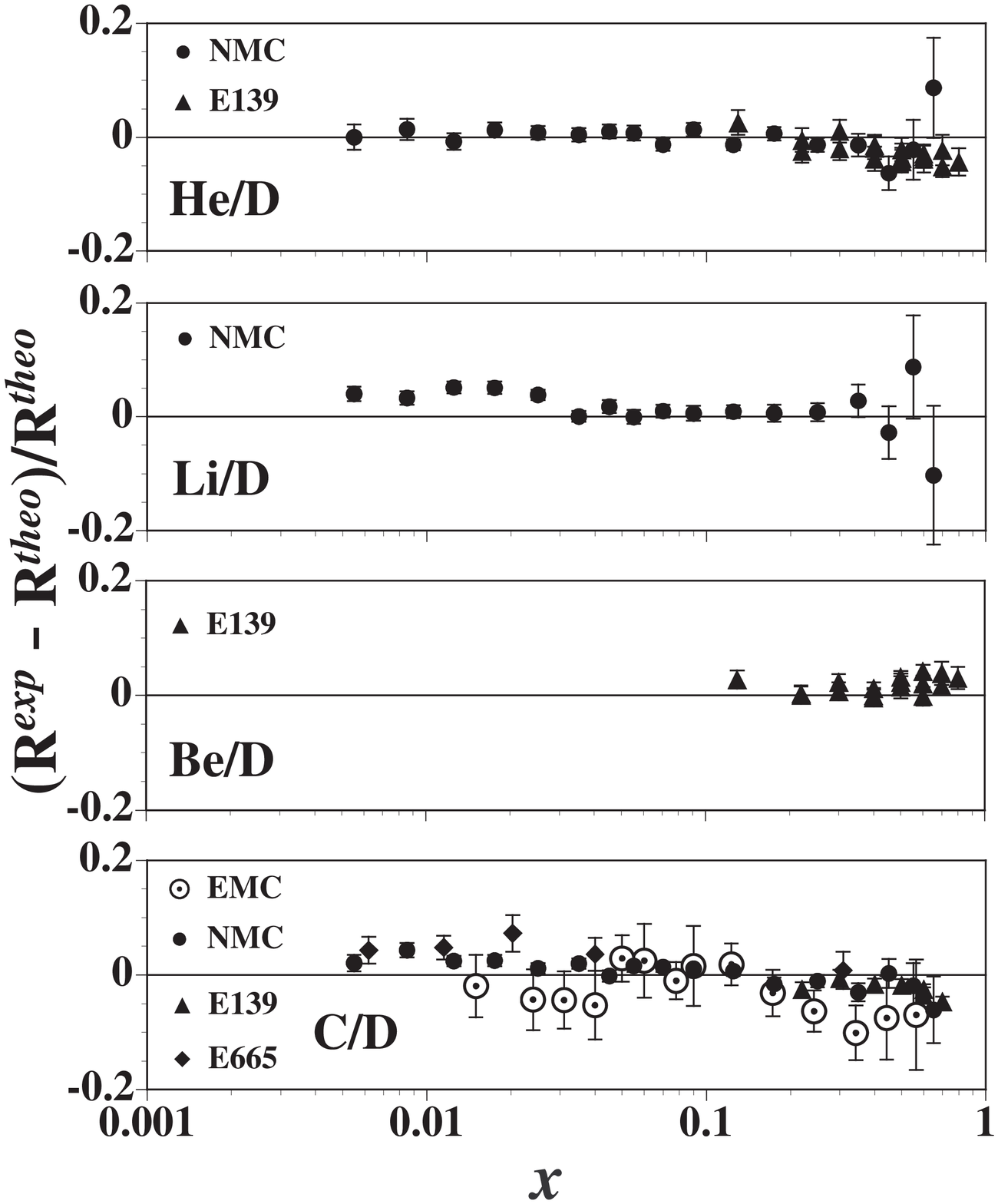}
\end{center}
\vspace{-0.7cm}
\vspace{0.0cm}
\begin{center}
     \includegraphics[width=0.30\textwidth]{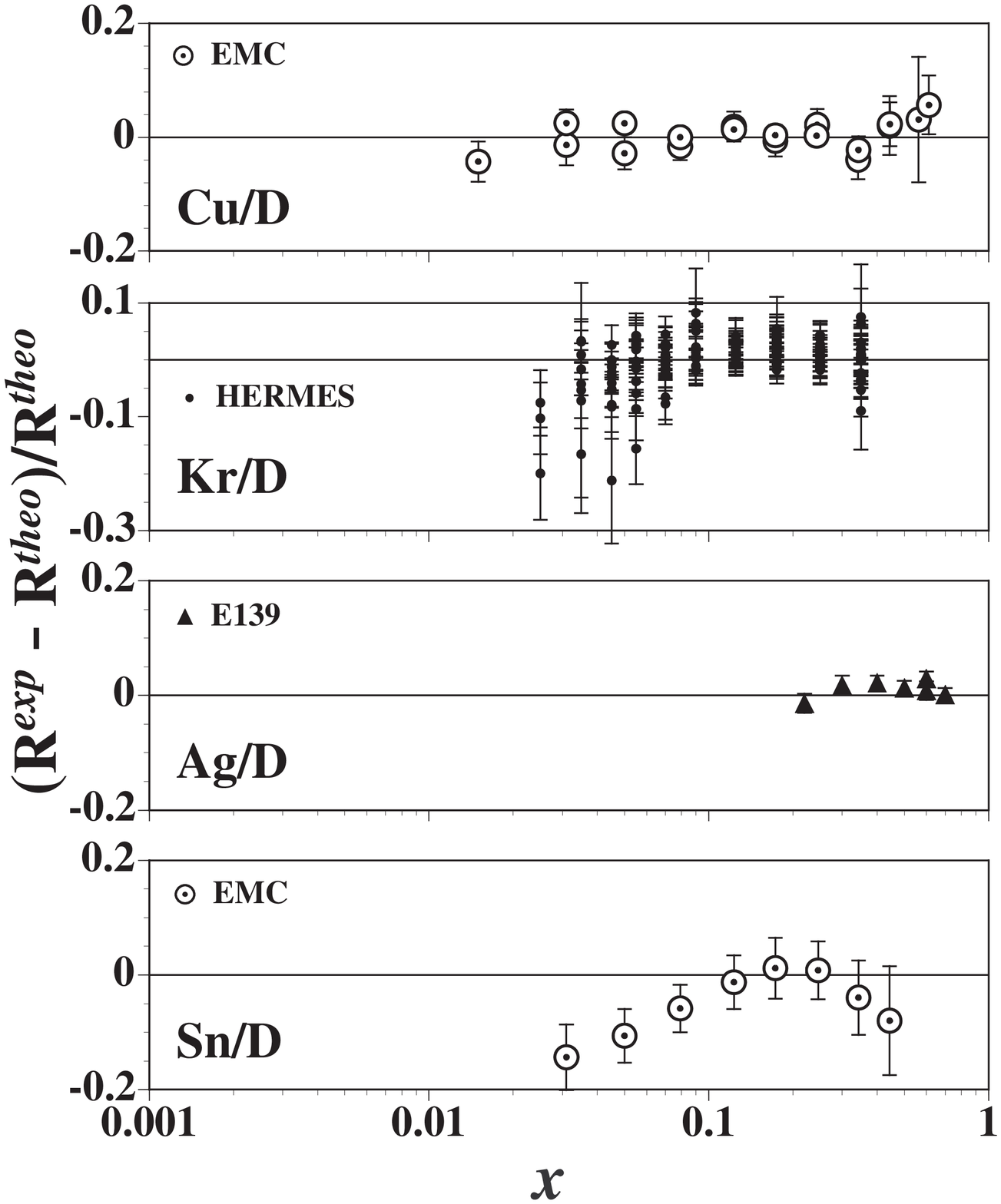}
\end{center}
\vspace{-0.7cm}
\vspace{0.0cm}
\begin{center}
     \includegraphics[width=0.30\textwidth]{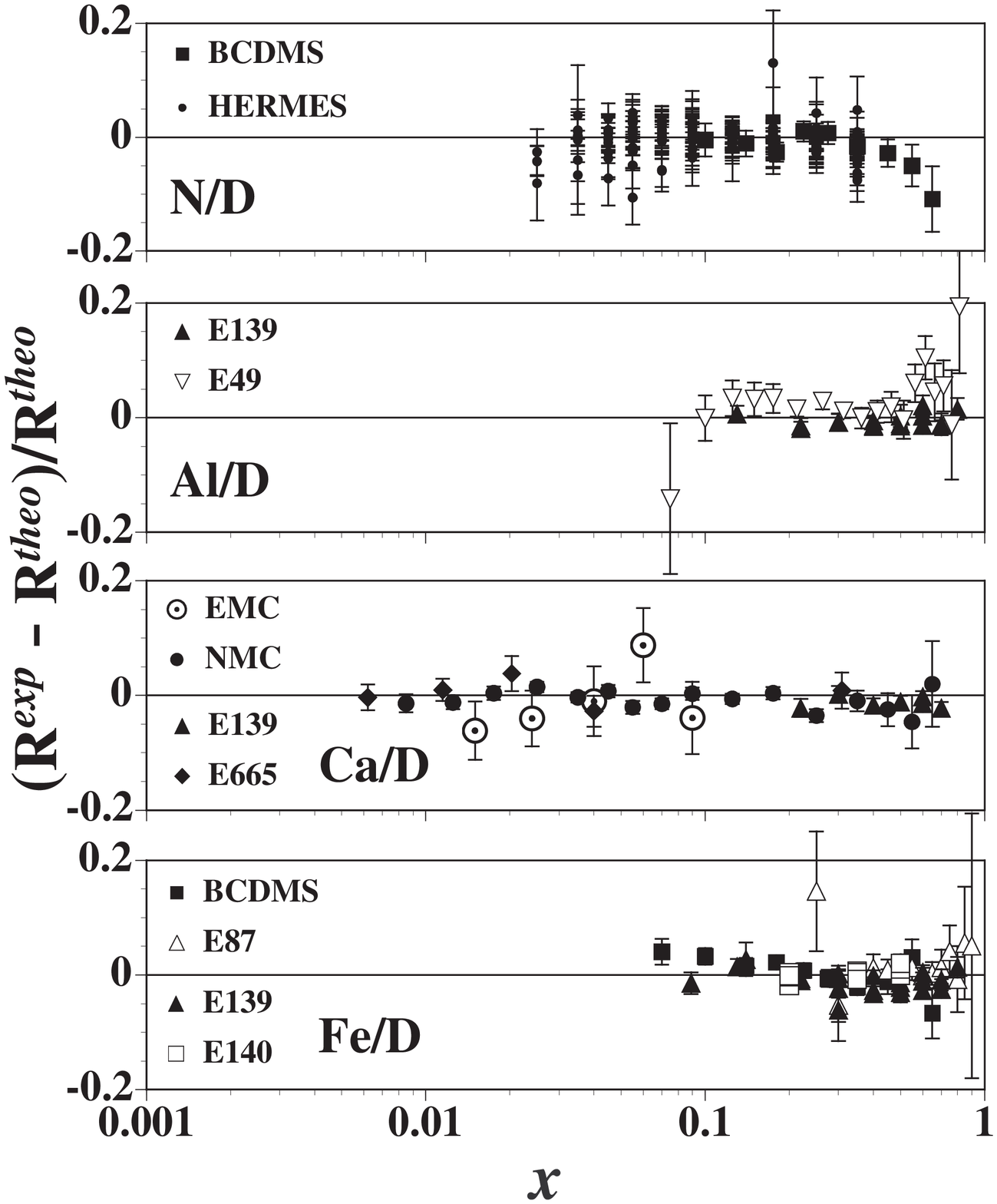}
\end{center}
\vspace{-1.2cm}
\caption{Fit results are compared with experimental data for $R=F_2^A/F_2^D$
         at the same experimental $Q^2$ points. The fractional differences
         $(R^{exp}-R^{theo})/R^{theo}$ are shown.}
\label{fig:f2a-d}
\end{figure}

The Drell-Yan data $\sigma_{DY}^{pCa}/\sigma_{DY}^{pD}$ are compared
with the fit result in Fig. \ref{fig:dycad}.
The parametrization curve and the uncertainty range are calculated
at $Q^2$=50 GeV$^2$, whereas the data are taken at various $Q^2$ points.
In the region $x<0.1$, the cross-section ratio is almost the same as
the antiquark ratio $\bar q^{Ca}/\bar q^{D}$, so that the data play
a role of fixing the nuclear antiquark distributions at $x \sim 0.1$.

The actual comparison with the experimental data should be done
at the same $Q^2$ points. In order to estimate the fit result, we show
the ratios $(R^{exp}-R^{theo})/R^{theo}$ in Fig. \ref{fig:f2a-d}.
Here, $R^{exp}$ indicates an experimental $F_2$ ratio
and $R^{theo}$ does a ratio by the parametrization. The theoretical
ratios are calculated at the experimental $Q^2$ points.
Among the used data, only $F_2^A/F_2^D$ type data are shown
in Fig. \ref{fig:f2a-d}. In general, the data are well explained by
the parametrization. However, there are slight deviations 
in small nuclei as indicated in Table \ref{tab:chi2}.
For example, the lithium data at small $x$
are underestimated. On the other hand, the tin data are overestimated
at small $x$. We need more complicated $A$ dependence for explaining
all the nuclei.

\subsection{$Q^2$ dependence}

\begin{figure}[b!]
\vspace{-0.5cm}
\begin{center}
     \includegraphics[width=0.42\textwidth]{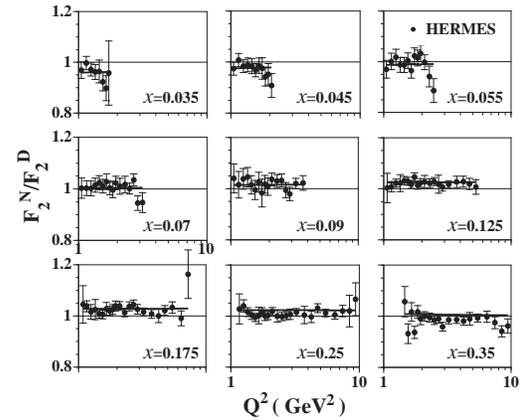}
\end{center}
\vspace{-1.2cm}
\caption{$Q^2$ dependence of $F_2^N/F_2^D$ is shown in comparison
          with the data. The curves indicate fit results.}
\vspace{-0.4cm}
\label{fig:q2dep}
\end{figure}

Next, $Q^2$ dependence of $F_2$ is calculated and it is shown with 
the $F_2^N/F_2^D$ data in Fig. \ref{fig:q2dep}, where $N$ indicates
nitrogen. The figure shows that the $Q^2$ dependence of the ratio
$F_2^N/F_2^D$ is not very obvious experimentally, which makes
the determination of the nuclear gluon distributions difficult.
The ratio tends to decrease with increasing $Q^2$ at $x=0.035$ and
$x=0.045$. There is a same tendency in the $F_2^{Kr}/F_2^D$ data
by the HERMES collaboration. However, the ratio $F_2^{Sn}/F_2^C$
increases with increasing $Q^2$ at $x=0.0125 \sim 0.045$
according to the NMC collaboration \cite{npdf04}. It seems that
the $Q^2$ variations are not consistent each other between the HERMES
and NMC data. Since such $Q^2$ dependence is crucial in determining
nuclear gluon distributions, we hope that the $Q^2$ dependence will be
accurately measured in the small-$x$ region.

\subsection{NPDFs with uncertainties}

\begin{figure}[b!]
\vspace{-0.5cm}
\begin{center}
     \includegraphics[width=0.28\textwidth]{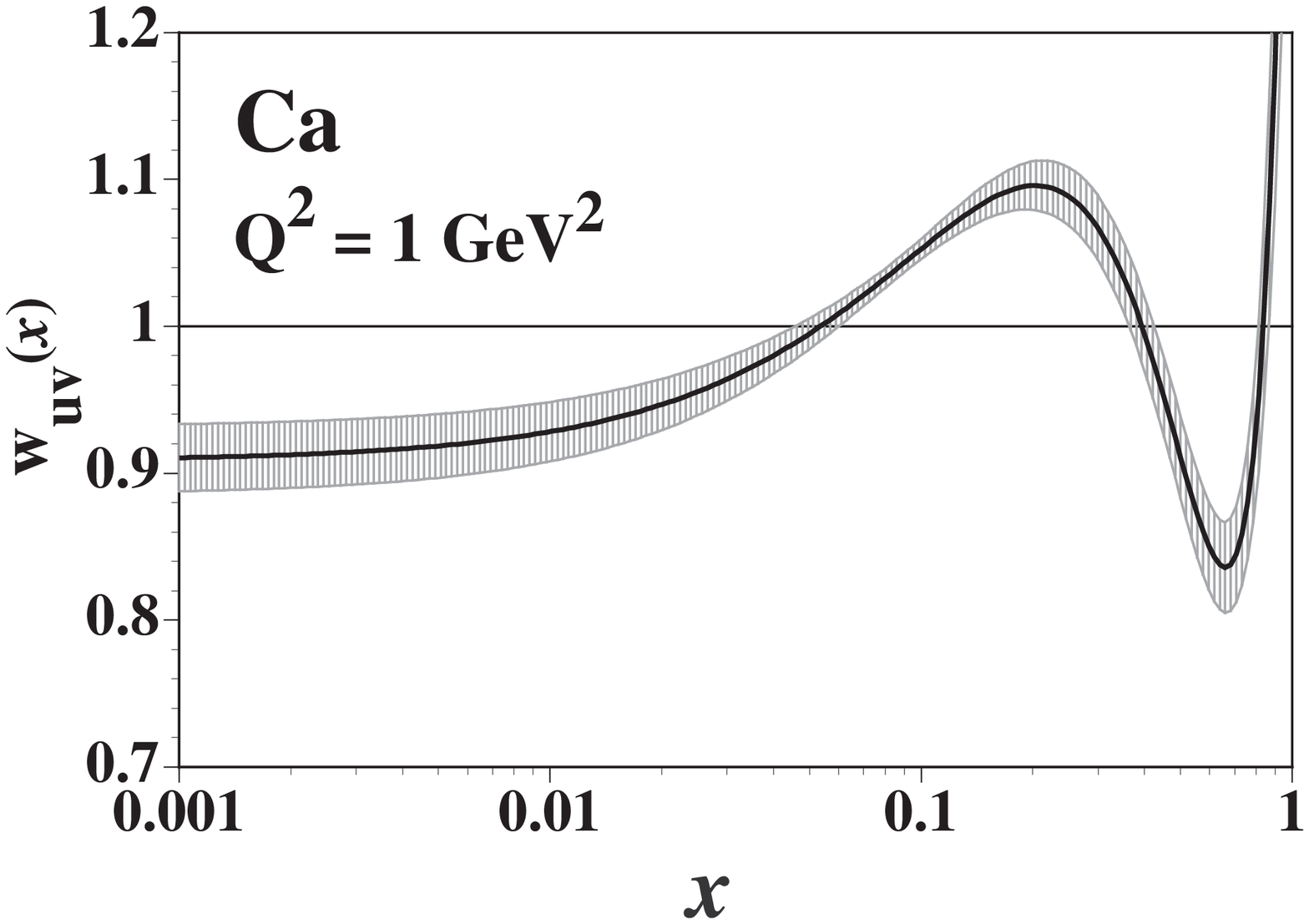}
\end{center}
\vspace{-0.7cm}
\vspace{0.0cm}
\begin{center}
     \includegraphics[width=0.28\textwidth]{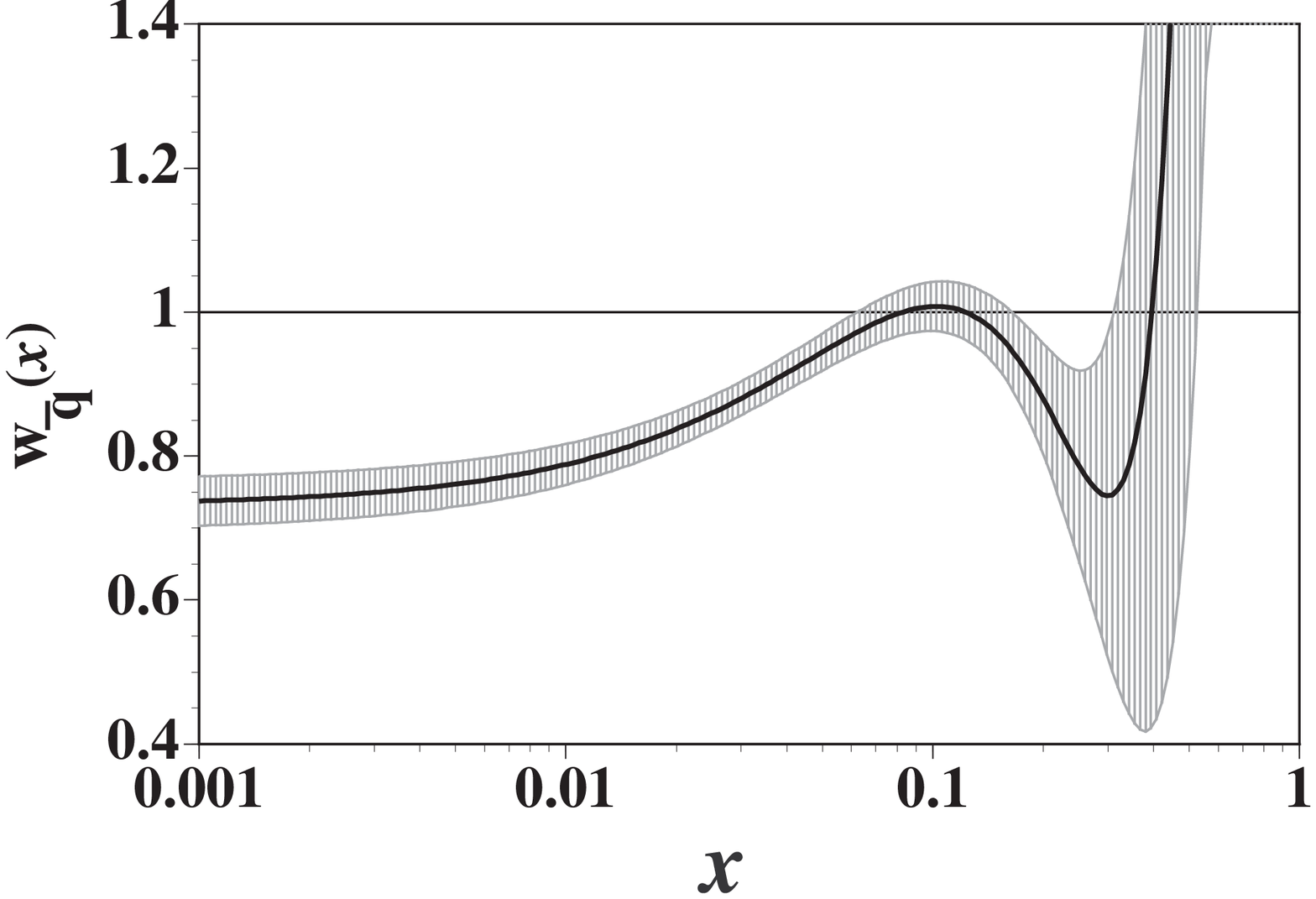}
\end{center}
\vspace{-0.7cm}
\vspace{0.0cm}
\begin{center}
     \includegraphics[width=0.28\textwidth]{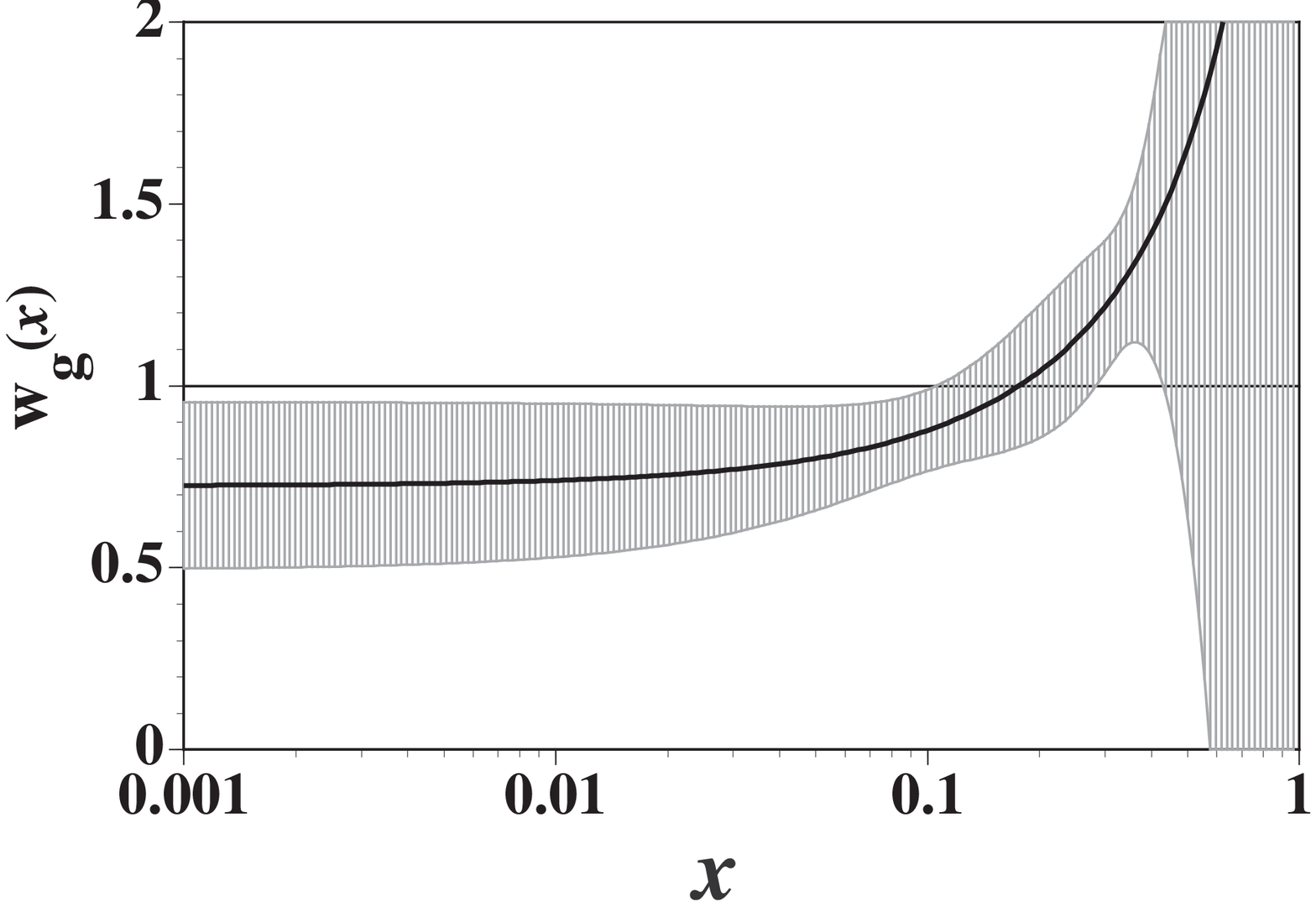}
\end{center}
\vspace{-1.2cm}
\caption{Nuclear modifications of valence-quark, antiquark, and gluon
         distributions in the calcium nucleus at $Q^2$=1 GeV$^2$.}
\vspace{-0.5cm}
\label{fig:wca}
\end{figure}

From the $\chi^2$ fit, we obtain the optimum NPDFs with the uncertainties.
As an example, the weight functions are shown for the calcium nucleus
in Fig. \ref{fig:wca}. These functions indicate nuclear modifications
of the valence-quark, antiquark, and gluon distributions in the calcium
at $Q^2$=1 GeV$^2$ by definition. The shaded areas indicate the uncertainties
estimated by the Hessian method.

The valence-quark distributions are well determined at medium $x$
by the $F_2^A$ data because the $F_2^A$ is dominated by them in
this $x$ region. In the small-$x$ region, the valence distributions
are fixed by baryon-number and charge conservations. 
These are the reasons why the uncertainties are rather small.
However, it is worth while investigating the valence-quark shadowing
at the NuMI and neutrino-factory projects \cite{numi-nufact}
despite the present result.

The antiquark distributions are determined at small $x$ because of
the shadowing data of $F_2^A$ at small $x$. They are also fixed 
at $x \sim 0.1$ by the Drell-Yan data. We notice that the uncertainties
are fairly small in these regions. However, they cannot be determined
at $x>0.2$ as indicated by the large uncertainties in Fig. \ref{fig:wca}. 
We need new experiments such as J-PARC and Fermilab-P906 
in this large-$x$ region \cite{dyexp}.

The gluon distributions cannot be fixed at this stage although they
seem to be shadowed at small $x$. In fact, the uncertainties are huge
in the whole $x$ region. The difficulty is mainly because
accurate scaling violation data are not available at small $x$
as shown in Fig. \ref{fig:q2dep}. The $Q^2$ variations
in the $x \sim 0.01$ region are not measured accurately, and
it is also obvious from Fig. \ref{fig:xq2} that
the small-$x$ ($x=10^{-3} \sim 10^{-4}$) data themselves do not exist.
In addition, a next-to-leading-order analysis could be important
for incorporating the gluon contributions into the parametrization.

From the analysis, we obtain the NPDFs from the deuteron to a heavy nucleus
with $A\sim 208$. We provide the NPDFs for general users by providing
a code, which is explained in the next section.
The NPDFs could be used for any high-energy nuclear reactions. 
In order to use them for the present neutrino oscillation experiments
in a medium-energy region, an extension from the DIS region to
the resonance one should be investigated. Such an effort has been done,
for example, in Ref. \cite{by02} for the structure function $F_2$
of the nucleon. Obviously, we need a similar study for extending
the NPDFs to the smaller $Q^2$ region in order to calculate
nuclear corrections to the cross sections in the neutrino oscillation
experiments.

\section{Code for calculating NPDFs}
\label{code}

A useful code could be obtained from the web site in Ref. \cite{npdf04}.
In the package (npdf04.tar.gz), grid data, npdf04 subroutine (npdf04.f),
and a sample file (sample.f) are provided. 
Running the subroutine, one obtains the NPDFs at given $x$ and $Q^2$
for the analyzed nuclei, D, $^4$He, Li, Be, C, N, Al, Ca, Fe, Cu,
Kr, Ag, Sn, Xe, W, Au, and Pb. 
The kinematical ranges for the provided NPDFs are $10^{-9} \le x \le 1$
and 1 GeV$^2 \le Q^2 \le 10^8$ GeV$^2$.
The smallest $x$ point of the data is $x_{min}$=0.0055, so that one should
note that the NPDFs at $x<x_{min}$ are not tested experimentally.
However, we provide the small-$x$ distributions in case that one
uses the NPDFs for integrating them over a wide range of $x$.
The maximum $Q^2$ of the data is 173 GeV$^2$. The $Q^2$
variations above this point are also not tested experimentally. 
We simply assumed the standard DGLAP $Q^2$ evolution equations
for extending the $Q^2$ region up to $Q^2=10^8$ GeV$^2$ for
feasibility studies of future experimental facilities.

For a nucleus other than the provided ones (D, He, $\cdot\cdot\cdot$, Pb),
the NPDFs could be calculated by following the procedure in Appendix of
Ref. \cite{npdf04}. Namely, the nuclear dependent parameters
$a_{u_v}$, $a_{d_v}$, $a_{g}$ are calculated by Eq. (A1) of
Ref. \cite{npdf04}, and other parameter values are taken from Table II.
Next, the NPDFs for a nucleus can be calculated at $Q^2$=1 GeV$^2$
by using the analytical expressions with the determined parameters.
Then, one may evolve the NPDFs to a given $Q^2$ point by
one's $Q^2$ evolution code or, if it is not available,
by the code in Ref. \cite{bf1}. The analyzed nuclei are up to $A=208$.
However, variations from the lead NPDFs to those
of the nuclear matter ($A\rightarrow \infty$) are small,
so that one may use the analysis results for large nuclei with $A>208$.

\section{Summary}
\label{summary}

We have investigated a parametrization of NPDFs by analyzing experimental
data on $F_2^A$ and Drell-Yan processes. In addition to obtaining the 
optimum NPDFs, we calculated their uncertainties by the Hessian method.
The antiquark distributions at small $x$ and valence-quark distributions
are well determined by the data. However, the antiquark distributions
at $x>0.2$ and the gluon distributions cannot be determined reliably.
We obviously need future experimental efforts for determining these
distributions. One could use the obtained NPDFs by running the provided
code.

\section*{Acknowledgments}

S.K. thanks Profs. F. Cavanna and M. Sakuda for taking care of
his participation in this workshop.  
He was supported by the Grant-in-Aid for Scientific Research from
the Japanese Ministry of Education, Culture, Sports, Science,
and Technology. 


\end{document}